\documentclass{PoS}

\usepackage{graphicx}
\usepackage{subcaption}
\usepackage{amsmath}
\usepackage{xspace}
\usepackage{booktabs}
\usepackage{multirow}
\usepackage{subcaption}
\usepackage{physics}

\DeclareRobustCommand{\NNLOJET}{\textsc{NNLOjet}\xspace}
\DeclareRobustCommand{\ensuremathrm}[1]{\ensuremath{\mathrm{#1}}\xspace}
\DeclareRobustCommand{\PZ}{{\ensuremathrm{Z}}\xspace}

\DeclareRobustCommand{\PH}{{\ensuremathrm{H}}\xspace}
\DeclareRobustCommand{\GeV}{{\ensuremathrm{GeV}}\xspace}
\DeclareRobustCommand{\rT}{\ensuremathrm{T}}
\DeclareRobustCommand{\alphas}{\ensuremath{\alpha_{\mathrm{s}}}\xspace}

\newcommand{\maple}{Maple\xspace}

\newcommand{\inimage}[1]{
    \raisebox{-0.5\height}{\includegraphics[height=0.5cm]{#1}} 
}

\newcommand{\mpm}{\ensuremath{\pm}}
\DeclareRobustCommand{\order}[1]{\ensuremath{\mathcal{O}\left(#1\right)}}

\newcommand{\parton}[1]{\ensuremathrm{#1}}
\newcommand{\Pg}{\parton{g}}
\newcommand{\Pq}{\parton{q}}
\newcommand{\PQ}{\parton{Q}}
\newcommand{\Pqb}{\parton{\bar{q}}}
\newcommand{\PQb}{\parton{\bar{Q}}}
\newcommand{\PR}{\parton{q'}}
\newcommand{\PRb}{\parton{\bar{q}'}}

\usepackage{color}
\usepackage{amssymb}
\definecolor{darkgreen}{rgb}{0.0, 0.6, 0.15}
\newcommand{\gct}{\color{darkgreen}\checkmark}
\newcommand{\rct}{\color{red}\checkmark}
\newcommand{\hct}{-}

\usepackage{tikz}
\usetikzlibrary{decorations.pathreplacing}
\newcommand{\tablebrace}[3]{
    \begin{tikzpicture}[remember picture, overlay]
        \draw [blue, decorate, decoration={brace, amplitude=4pt, raise=1pt}]
            (#2,#3)--(#2,0.37) node [blue,left=5pt, midway] {#1} ;
    \end{tikzpicture}
}
\newcommand{\figbrace}[3]{
    \begin{tikzpicture}[remember picture, overlay]
        \draw [black, decorate, decoration={brace, amplitude=4pt, raise=1pt}]
            (#3,#2)--(#3,#2+1.0) node [black,left=5pt, midway] {#1} ;
    \end{tikzpicture}
}
\definecolor{ballblue}{rgb}{0.0, 0.50, 0.89}
\newcommand{\flipfigbrace}[3]{
    \begin{tikzpicture}[remember picture, overlay]
        \draw [black, decorate, decoration={brace, amplitude=4pt, raise=1pt}]
            (#3,#2)--(#3,#2-1.0) node [ballblue,right=5pt, midway] {#1} ;
    \end{tikzpicture}
}

\title{ NNLO corrections to VBF Higgs boson production }

\ShortTitle{VBF Higgs at NNLO}

\author{\speaker{J.M. Cruz-Martinez}, E.W.N.\ Glover\\
        Institute for Particle Physics Phenomenology, Durham University, Durham, DH1 3LE, UK
    }
\author{ T.\ Gehrmann \\ 
    Physik-Institut, Universit\"at Z\"urich, Winterthurerstrasse 190,CH-8057 Z\"urich, Switzerland
}
\author{ A.~Huss \\\
     Theoretical Physics Department, CERN, CH-1211Geneva 23, Switzerland
}

\abstract{
    This talk expands on recently published results for the factorising next-to-next-to-leading order (NNLO) QCD corrections to Higgs boson production in the vector boson fusion (VBF) channel~\cite{Cruz-Martinez:2018rod}.
   The calculation is fully differential in the kinematics of the Higgs boson and the final state jets and is implemented in the \NNLOJET framework for computing higher-order QCD corrections. 
    We find the NNLO corrections to be limited in magnitude to about \mpm 5\% with a weak kinematical dependence in the transverse momenta and rapidity separation of the two tagging jets.
}

\FullConference{Loops and Legs in Quantum Field Theory (LL2018)\\
		29 April 2018 - 04 May 2018\\
		St. Goar, Germany}

\begin{document}

\section{Introduction}
The precise study of the properties of the Higgs boson has been at the core of the research program of the CERN Large Hadron Collider (LHC) since the announcement of its discovery on July 4th, 2012~\cite{Aad:2012tfa,Chatrchyan:2012xdj}.

At LHC energies, the Higgs boson can be produced via several mechanisms. The vector boson fusion (VBF) process is the
second-largest inclusive production mode, amounting to about 10\% of the dominant 
gluon fusion process. The detailed experimental study of the VBF production mode probes the 
electroweak coupling structure of the Higgs boson, thereby testing the Higgs mechanism of electroweak 
symmetry breaking.
It is therefore paramount to be able to identify VBF events from the background from other production modes.
Experimentally, this discrimination is achieved by exploiting the distinctive VBF signature: a Higgs boson in association with two jets which are strongly separated in rapidity and forming a dijet system of high invariant mass.

Perturbative corrections to Higgs boson production in VBF have been derived at next-to-leading order (NLO) in QCD~\cite{Figy:2003nv,Berger:2004pca,Figy:2004pt,Arnold:2008rz} and in the electroweak theory~\cite{Ciccolini:2007ec}. 
NNLO QCD corrections to the inclusive VBF Higgs production cross section were found to be very small~\cite{Bolzoni:2010xr}, 
which is confirmed by third-order (N3LO) corrections~\cite{Dreyer:2016oyx}.
More sizeable effects for fiducial cross sections were found in~\cite{Cacciari:2015jma}, using the inclusive results from~\cite{Bolzoni:2010xr} and the NLO QCD VBF Higgs-plus-three-jet result of~\cite{Figy:2007kv} in order to access the kinematical information of the jets and employing the projection-to-Born method.

Our implementation of this process at NNLO QCD in \NNLOJET~\cite{Cruz-Martinez:2018rod, Gehrmann:2018szu} using antenna subtraction is fully differential in all kinematical variables and allows us to compute any IR-safe observable at this order. We find discrepancies with the results originally published in~\cite{Cacciari:2015jma}.
These discrepancies are traced back to an error in the VBF Higgs-plus-three-jet calculation of~\cite{Figy:2007kv} upon which~\cite{Cacciari:2015jma} relies.
We find excellent agreement with the revised results.

    \subsection{VBF cuts}
    \label{sec:vbf_cuts}
    It is possible to devise a set of cuts which single out the VBF contribution by exploiting the leading-order (LO) topology of the process~\cite{Barger:1994zq, Rainwater:1998kj}: two incoming quarks deflected by the emission of weak vector bosons which fuse into the Higgs boson (see Fig.~\ref{fig:born}). Both deflected quarks become energetic jets at large opposite rapidities. Furthermore, the lack of colour exchange between the two initial state partons means activity in the central region is suppressed with respect to other Higgs production channels. Besides enhancing the relative contribution of VBF-like processes, these cuts also suppress interference effects between the two quark currents (present for identical quark flavours, e.g. $uu$ scattering).


    We select events in which the two tagging jets (defined as the two hardest jets found by the jet algorithm) are encountered in different hemispheres, with a rapidity gap of $\Delta y_{jj} \geq 4.5$ between them. A cut on the invariant mass of the two tagging jets of $M_{jj} > 600~\GeV$ further suppresses $s$-channel contributions~\cite{deFlorian:2016spz}.
Jets are defined using the anti-kt algorithm~\cite{Cacciari:2008gp} with a radius parameter $R = 0.4$ and are required to have a transverse momentum greater than 25~\GeV and a rapidity of less than 4.5 which ensures both tagging jets are found in opposite hemispheres.


\section{Implementation}
\label{sec:implementation}

\subsection{VBF process}
\label{sec:vbf_process}

\begin{figure}[t]
  \centering
  \figbrace{$J^{(0)}_{\mu}(\Pq,\Pq')$}{2.1}{0.75}
  \figbrace{$J^{(0)}_{\nu}(\PQ,\PQ')$}{0.2}{0.45}
  \flipfigbrace{$\text{M}_{\text{\tiny VVH}}^{\mu\nu}$}{2.16}{4.8}
  \includegraphics[width=0.3\linewidth]{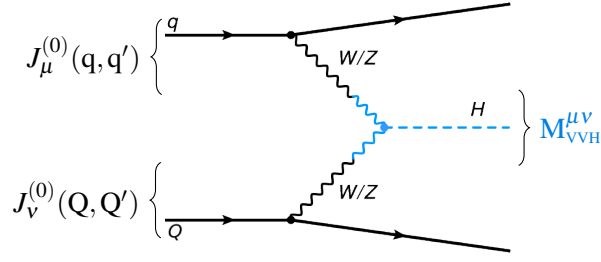}\hfill
  \caption{Born-level vector boson fusion process.\label{fig:born}}
\end{figure}

The VBF production mode can be seen as two independent Deep Inelastic Scattering (DIS) processes where the two off-shell vector bosons fuse through a VVH vertex, as depicted in Fig.~\ref{fig:born}. In this picture, the VBF process is formed by two quark currents, $J^{(l)}_{\mu}(\Pq,\Pq',\ldots)$ and $J^{(l)}_{\nu}(\PQ,\PQ',\ldots)$, connected through a weak boson-Higgs vertex, $\text{M}_{\text{\tiny VVH}}^{\mu\nu}$. Explicitly, the Born-level amplitude can be written as:
\begin{equation}
    \mathcal{M}^{(0)}\left(\Pq,\PQ,\PQ',\Pq'\right) = J^{(0)}_{\mu}(\Pq,\Pq') \ \text{M}_{\text{\tiny VVH}}^{\mu\nu} \ J^{(0)}_{\nu}(\PQ,\PQ') \label{eq:baseamp},
\end{equation}
where the labels \Pq and \PQ refer to a quark or antiquark of any massless flavour.\footnote{We consider five massless flavours: u, d, c, s, b}

At LO, the only matrix element that enters the calculation is the square of Eq.~\eqref{eq:baseamp}:
\begin{equation}
    \text{C}_{0g}^{(0)} = |\mathcal{M}^{(0)}\left(\Pq,\PQ,\PQ',\Pq'\right)|^{2}, \label{eq:ctype}
\end{equation} 
i.e., we neglect interference effects from the special case in which the flavours of the two quark currents \Pq and \PQ coincide and the two quarks are indistinguishable:
\begin{equation}
    \text{D}_{0g}^{(0)} = \frac{2}{N}\Re\left\{\mathcal{M}^{(0)}\left(\Pq,\PQ,\PQ',\Pq'\right)\mathcal{M}^{(0)}\left(\Pq,\PQ,\Pq',\PQ'\right)^{*}\right\}. \label{eq:dtype}
\end{equation}
These matrix elements account for interference effects between the two currents and are suppressed by a factor of $\frac{1}{N}$ and kinematically in the regions of the phase space defined by the VBF cuts.

Similarly, we neglect the color and kinematically suppressed contributions due to the interference of gluons radiated from different currents~\cite{Ciccolini:2007ec,Campanario:2013fsa,Bolzoni:2011cu} at higher orders.
Interference effects between VBF and other Higgs plus-two-jets production channels have also been shown to be negligible~\cite{Andersen:2007mp}.

\begin{figure}[t] 
  \centering 
  \includegraphics[width=0.30\linewidth]{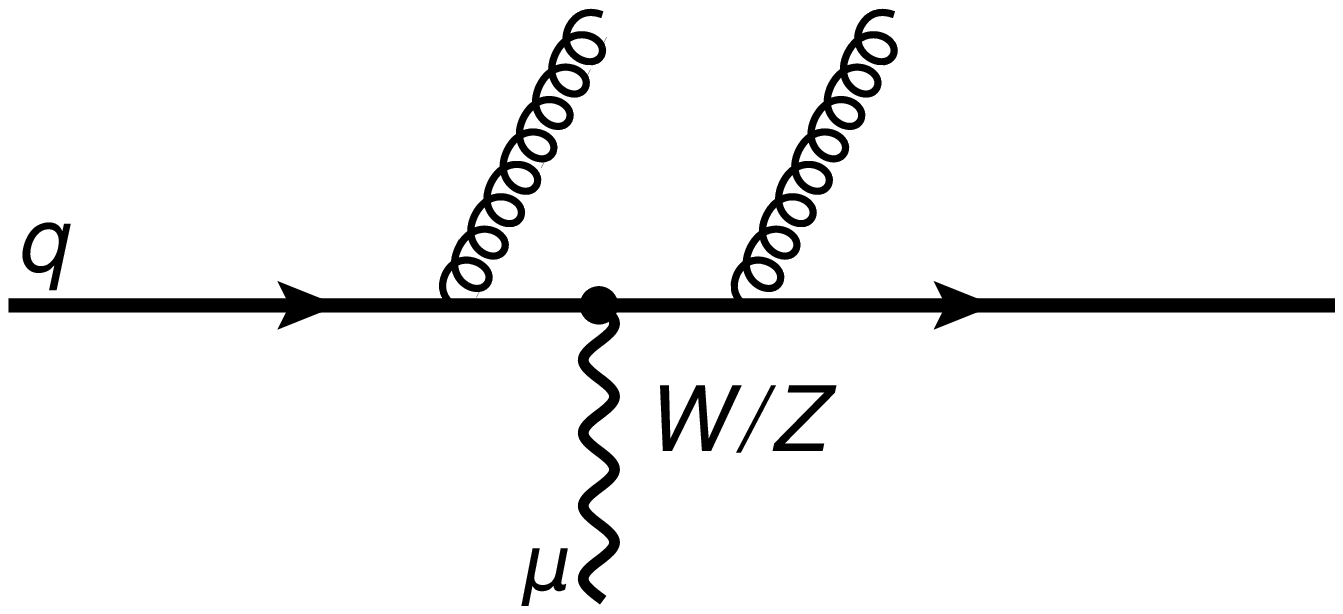}\hfill
  \includegraphics[width=0.30\linewidth]{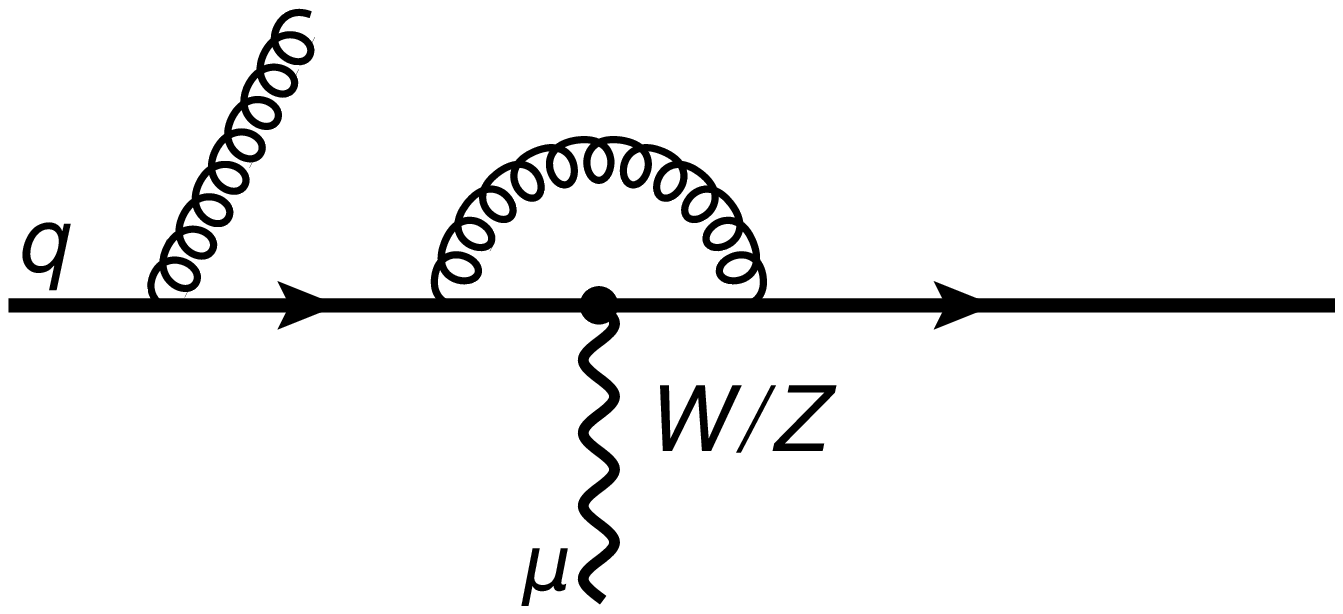}\hfill
  \includegraphics[width=0.30\linewidth]{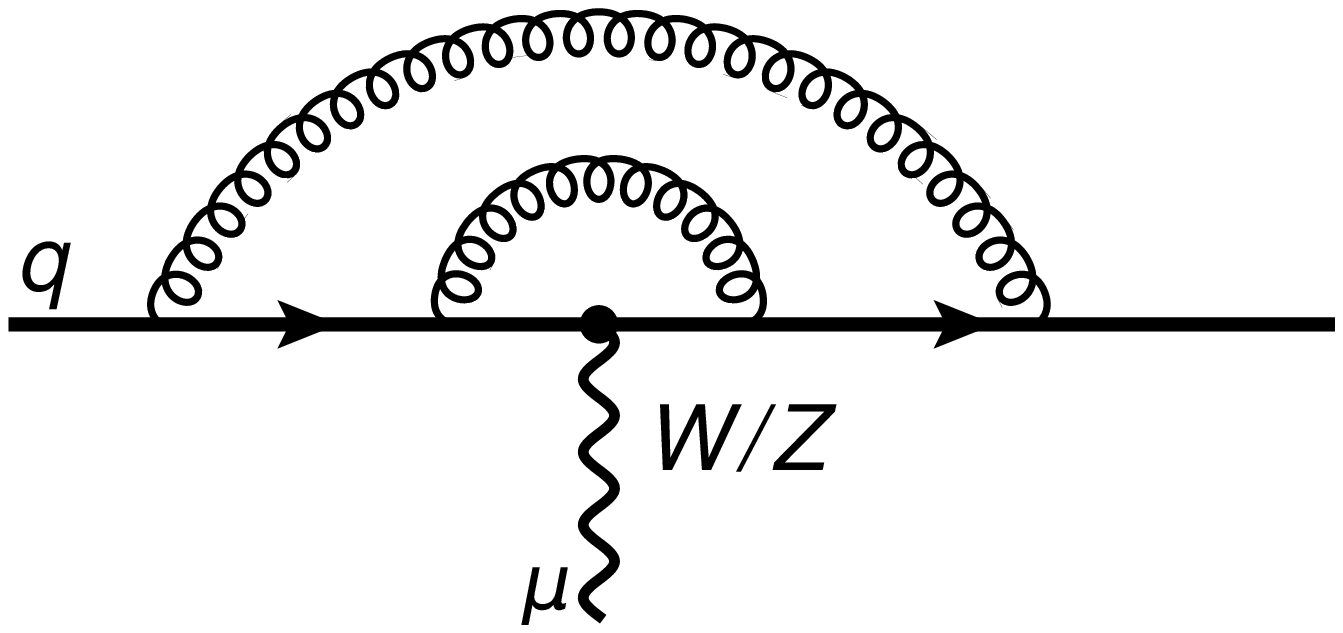}
  \caption{Examples of second order QCD corrections (RR, RV, VV) to the quark currents.\label{fig:nnlofeyn}}
\end{figure}

Therefore the subset of diagrams we keep is equivalent to the so-called ``structure function approach''~\cite{Han:1992hr}, as we are effectively considering QCD corrections to each current separately (Fig.~\ref{fig:nnlofeyn}).

\begin{table}[t]
    \footnotesize
    \centering
     \begin{tabular}{ c | l }
         \toprule
         Level & \multicolumn{1}{c}{ Processes } \\
         \midrule
         LO, V, VV & {\Pq\PQ $\to$ \Pq\PQ\PH}  \\
         \midrule
         R, RV & \Pq\PQ $\to$ \Pq\PQ\Pg\PH ; \Pq\Pg $\to$ \Pq\PQ\PQb\PH \\
         \midrule
         \multirow{3}{*}{RR} &  \Pq\PQ $\to$ \Pq\PQ\Pg\Pg\PH; \Pq\PQ $\to$ \Pq\PQ\PR\PRb\PH  \\
                                       & \Pq\Pg $\to$ \Pq\PQ\PQb\Pg\PH ; \Pg\Pg $\to$ \Pq\Pqb\PQ\PQb\PH \\
         \bottomrule
    \end{tabular}
    \caption{Subprocesses that contribute to VBF-$2j$ at NNLO in \NNLOJET.\label{table:subproc}}
\end{table}

Table~\ref{table:subproc} lists all Higgs production subprocesses contributing up to \order{\alphas^2} for VBF-$2j$, VBF-$3j$ and VBF-$4j$. The labels \PR and \PRb in this table refer to the quarks produced when a gluon radiated from either of the currents subsequently splits into a quark-antiquark pair.

\subsection{\NNLOJET}

QCD NNLO corrections listed in Table~\ref{table:subproc} include contributions from double real radiation (RR), single real radiation at one loop (RV) and two-loop virtual (VV),
which introduce implicit infrared (IR) singularities (upon phase space integration) and explicit IR poles (from loop integrals) which need to be regulated. 
In order to render each contribution finite and numerically integrable we use the antenna subtraction technique~\cite{GehrmannDeRidder:2005cm,GehrmannDeRidder:2005aw,GehrmannDeRidder:2005hi,Daleo:2006xa} for the subtraction of real radiation singularities and the reintroduction of its integrated counterparts:
\newcommand{\dsigma}{{\rm d}\hat\sigma_{ij}}
    \begin{align}
\dsigma^{\text{NNLO}} & = 
\int_{\Phi_5}\dsigma^{RR} 
+ \int_{\Phi_4}\dsigma^{RV} 
+ \int_{\Phi_3}\dsigma^{VV} \nonumber \\ 
                      & = 
\int_{\Phi_5}\left(\dsigma^{RR} - \dsigma^{S}\right)
+ \int_{\Phi_4}\left(\dsigma^{RV} - \dsigma^{T}\right)
+ \int_{\Phi_3}\left(\dsigma^{VV} - \dsigma^{U}\right), \label{eq:antenna_total}
    \end{align}
where the integration is performed over the additional phase space of the radiated partons.

    The numerical parton-level implementation is performed in the Fortran-based \NNLOJET framework~\cite{Gehrmann:2018szu}, which provides the phase-space generator, event handling and analysis routines as well as subroutines for all unintegrated and integrated antenna functions used to construct the subtraction terms.                                                         
    
Process-dependent Fortran code for the various subtraction terms (which need to be manually assembled in \maple), different parton orderings and prefactors stemming from symmetries and colour configurations is autogenerated using \maple.  

    For our numerical computations, we use LHAPDF~\cite{Buckley:2014ana} with the NNPDF3.0 NNLO parton distribution functions~\cite{Ball:2014uwa} with the value of $\alphas(M_{\PZ})=0.118$, and $M_{\PH}= 125~\GeV$, which is compatible with the combined results of ATLAS and CMS~\cite{Aad:2015zhl}. 

We choose as central value for the renormalisation and factorisation scales:
\begin{equation}
    \mu_0^{2} (p_{\rT}^{\PH}) = \frac{M_{\PH}}{2}\sqrt{\left(\frac{M_{\PH}}{2}\right)^2 + \left(p_{\rT}^{\PH}\right)^2}.
    \label{eq:scale_choice}
\end{equation}

\subsection{Validation}
\label{sec:validation}

\begin{table}[t]
    \footnotesize
    \centering
     \begin{tabular}{ r |  l*{7}{c   }}
\toprule
     Level & ME & Spikes & $\frac{1}{\epsilon}$ & Layer & Scale & Phase Space & Inclusive \\
     \midrule \tablebrace{LO}{-0.5}{-0.15} 
     B & \rct & \hct & \hct & \hct & \hct & \gct & \rct \\
     \midrule \tablebrace{NLO}{-0.5}{-0.6}
     R  & \rct & \gct & \hct & \gct & \gct & \gct & \rct \\
     V  & \rct & \hct & \gct & \gct & \gct & \gct & \rct \\
     \midrule \tablebrace{NNLO}{-0.2}{-1.05}
     RR & \rct & \gct & \hct & \gct & \gct & \gct & \rct \\
     RV & \rct & \gct & \gct & \gct & \gct & \gct & \rct \\
     VV & \hct & \hct & \gct & \gct & \gct & \gct & \rct \\
\bottomrule
    \end{tabular}
    \caption{Red ticks refer to tests against external tools, green ticks are internal \NNLOJET tests. Non-applicable tests are marked with an hyphen. \label{table:test_summary}}
\end{table}

Computations of NNLO QCD corrections are highly non-trivial as many different ingredients need to be considered and, crucially, tested.
Table~\ref{table:test_summary} presents a summary of the test suite we used for our calculation.
These include internal consistency checks within the \NNLOJET framework as well as checks against well established external tools.

The first test in Table~\ref{table:test_summary}, labeled ME, is a validation of our independent implementation of the matrix elements against automated tools such as Madgraph~\cite{Alwall:2014hca} and OpenLoops~\cite{Cascioli:2011va}. Once we correct for the difference in parameters and conventions, we find machine precision agreement with both programs for all parton-level configurations for any given phase space point. Some examples are given in Table~\ref{table:pointwise}.
\begin{table}[t]
    \centering
    \scriptsize
    \begin{tabular}{c c c c}
        \toprule
        Process                                & \NNLOJET                        & Madgraph / OpenLoops           & Ratio \\  
        \midrule
        \inimage{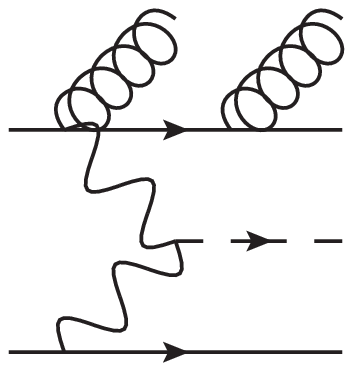}     & {\tt 5.1406085025982153E-014}  & {\tt 5.1406085025982185E-014}  & {\tt 1.0000000000000007} \\ \\
        \inimage{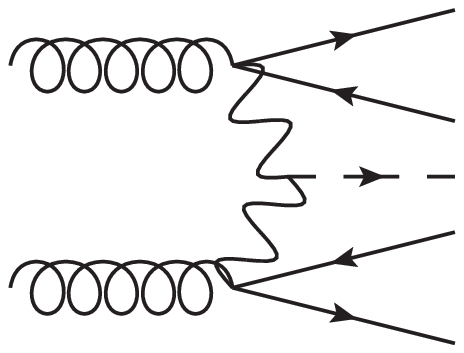}  & {\tt 4.0982703031871614E-017}  & {\tt 4.0982703031871552E-017}  & {\tt 1.00000000000000163}  \\ \\
        \inimage{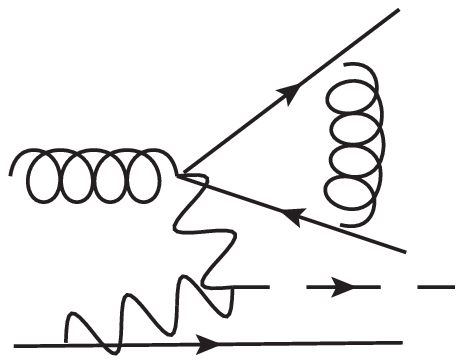} & {\tt -5.9293779081794126E-011} & {\tt -5.9293779081799606E-011} & {\tt 0.99999999999990763} \\ \\
        \bottomrule
    \end{tabular}
    \caption{Pointwise comparison against reference tools for a selection of RR and RV matrix elements that enter the calculation. All tests have been performed for all colour levels and possible configurations of initial and final states allowed by the approach described in Section~\ref{sec:vbf_process}. \label{table:pointwise}}
\end{table}

\begin{figure}[t]
  \centering
  \includegraphics[width=0.40\linewidth]{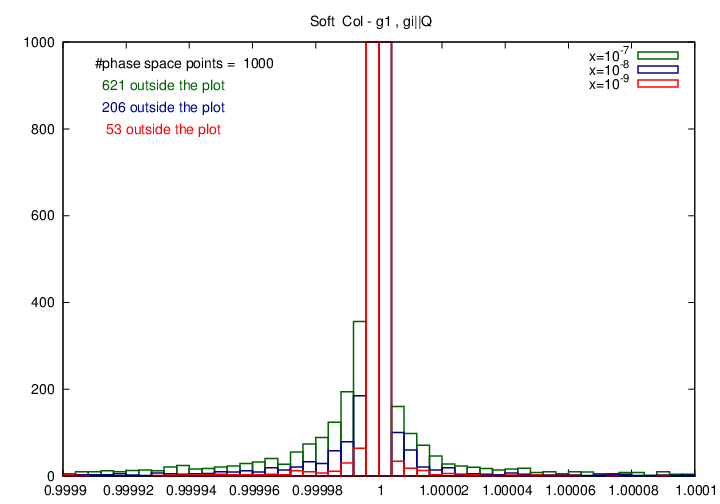}\qquad
  \includegraphics[width=0.40\linewidth]{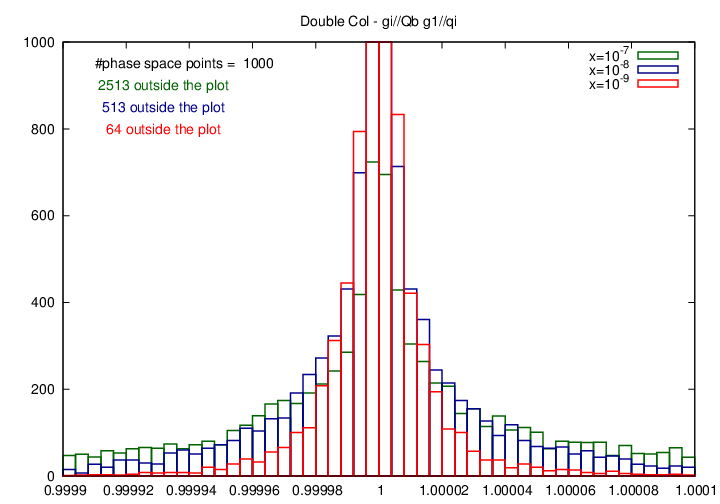}\qquad
  \caption{``Spike plots'' generated for two different singular limits. As we approach the unresolved region by reducing the value of $x<\frac{s_{ij}}{\hat{s}}$, the cancellation becomes more precise.\label{fig:spike}}
\end{figure}

Matrix elements with extra radiation or loops introduce singularities which need to be cancelled by subtraction terms. 
We ascertain the cancellation of the implicit singularities of the process by means of frequency histograms of the ratio of matrix element and subtraction term for different phase space points chosen randomly around the unresolved limits.
Fig.~\ref{fig:spike} shows some examples of these ``spike plots'' which demonstrate that the subtraction term successfully approximates the matrix element in the unresolved regions.

Virtual matrix elements, on the other hand, posses explicit $\frac{1}{\epsilon}$ infrared poles arising from loop integration which must be cancelled by the integrated counterterms. This is tested analytically by the autogeneration routines. Taken together, these two tests ensure that the cross section is finite and integrable over the entire range of the phase space.

    Once matrix elements and subtraction terms are properly validated, the final step is to check that Eq.~\eqref{eq:antenna_total} still holds, meaning the total value of the cross section after adding together all different ``layers'' (RR, RV and VV) remains unchanged. This is equivalent to verifying our subtraction terms fulfill the equality:
\begin{equation}
        \int_{\Phi_5} \dsigma^{S} + \int_{\Phi_4} \dsigma^{T} + \int_{\Phi_3} \dsigma^{U} = 0\label{eq:layer},
    \end{equation}
i.e., that all antenna counterterms subtracted in $\dsigma^{S}$ and $\dsigma^{T}$ (cancelling the implicit singularities of the RR and RV contributions) are reintroduced back as integrated antennae in the same amount in $\dsigma^{T}$ and $\dsigma^{U}$ (capturing the explicit poles). This is shown graphically in Fig.~3 of Ref.~\cite{Currie:2013vh}.

    We exploit the autogenerated nature of the \NNLOJET framework by reading in the \maple files that are used to autogenerate the Fortran code and checking that Eq.~\eqref{eq:layer} holds for every colour level for every parton ordering.

    By themselves, these four tests are quite powerful in asserting the validity of the results. However, there are other details that might escape the scope of these tests and they can be captured by further checks done at the integration level. For example we can apply the Renormalisation Group Equation to an observable for a fixed value of the renormalisation and factorisation scale and directly verifying the same result is obtained numerically in \NNLOJET for different scale choices.
    
    The high dimensionality of the phase space for processes like VBF (five particles in the final state at NNLO) also puts a heavy burden in the numerical code and so it is necessary to write a phase space which adapts to the topology of the process. It is also a requirement to verify that the entire range of the phase space is being probed by the integrator, specially in the singular regions. 
    We can do this by lifting all cuts in the final state partons. Table~\ref{table:fully_inclusive} shows that we find excellent agreement with the results of Ref.~\cite{Cacciari:2015jma} for the total inclusive cross section which was computed with the method of Ref.~\cite{Bolzoni:2010xr}. 
    In both these references the NNLO cross section is obtained directly from the DIS coefficient functions~\cite{Zijlstra:1992qd} which analytically combine together all parton level subprocesses of different state multiplicity.
    In contrast, we evaluate all subprocesses numerically using the antenna subtraction counterterms to regulate implicit singularities.
    The inclusive cross section is assembled from the sum of all different combinations considered in Section~\ref{sec:vbf_process}. 
   This is a highly non-trivial test of our implementation of all individual subprocesses and antenna counterterms.



\section{Results}
\label{sec:results}

In Table~\ref{table:fully_inclusive} we see that both NLO and NNLO contributions individually correspond to less than 3\% of the full inclusive cross section. In contrast
Table~\ref{table:fiducial} lists the total fiducial cross sections with VBF cuts. The VBF cuts increase the size of the NLO and NNLO effects, which are up to three times larger than for the inclusive configuration.

\begin{figure}[t]
    \centering
  \includegraphics[width=0.40\linewidth]{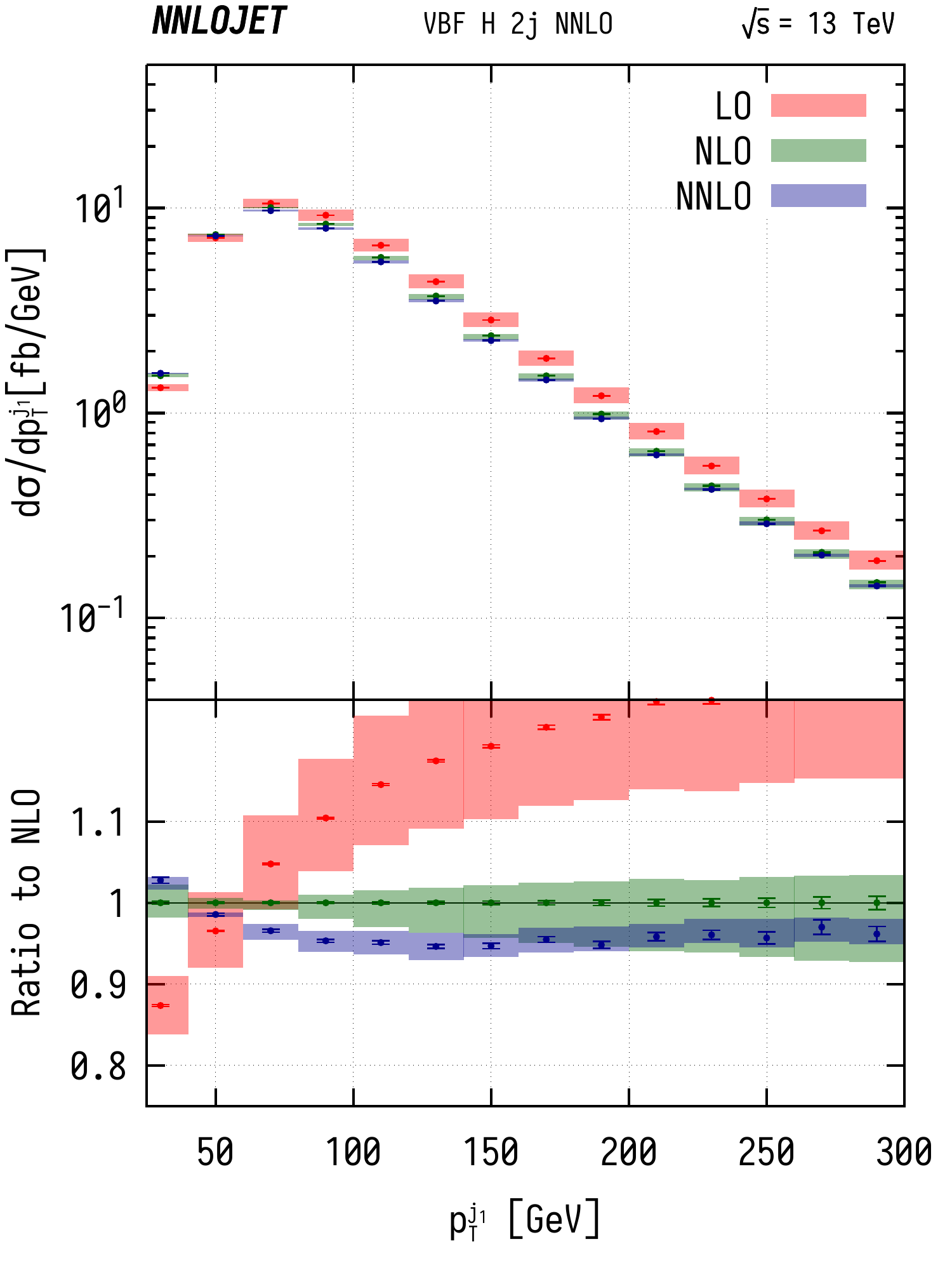}\qquad
  \includegraphics[width=0.40\linewidth]{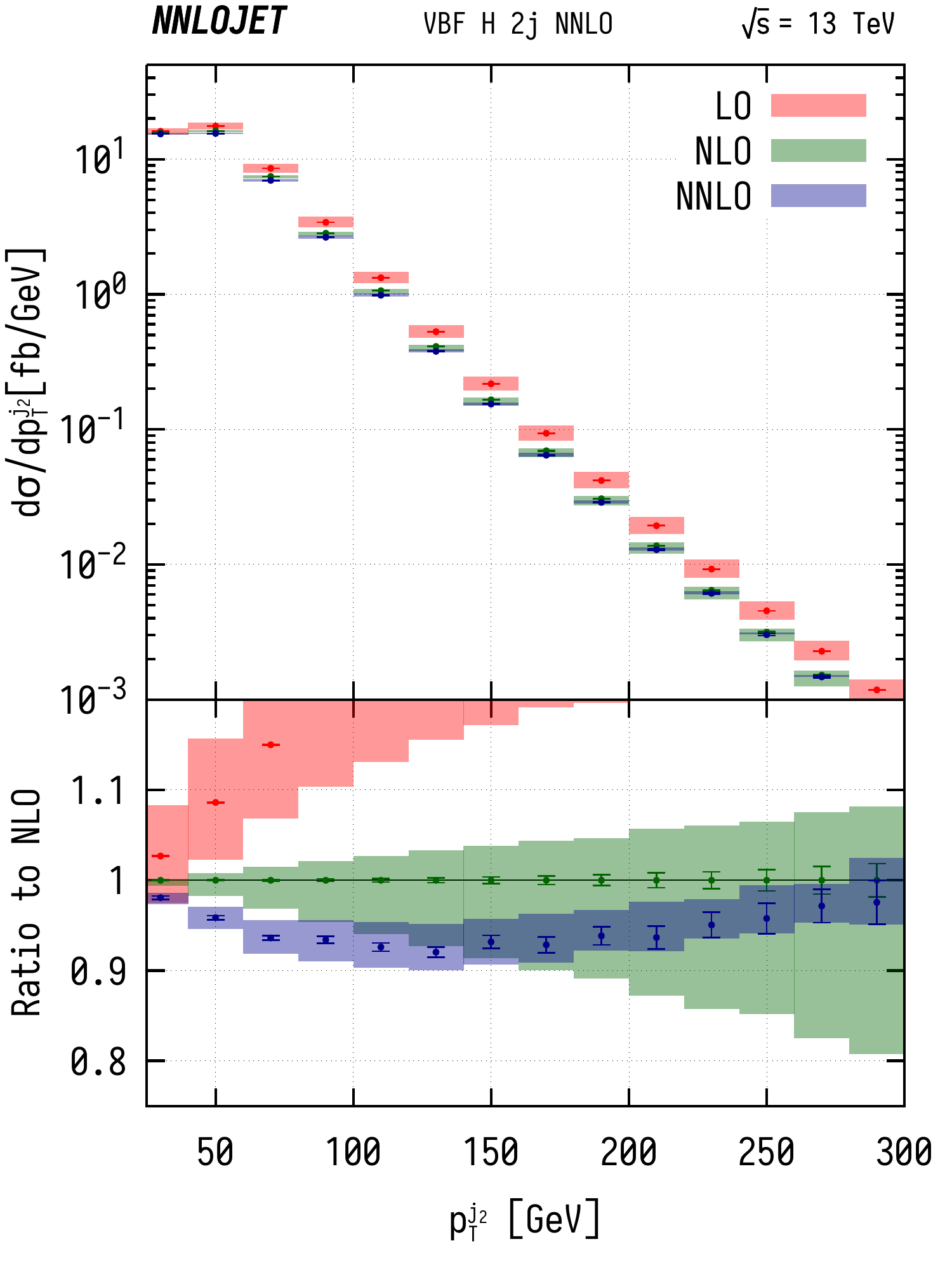}
  \caption{Transverse momentum distribution of the leading and subleading jet in VBF process.\label{fig:ptj}}
\end{figure}

The transverse momentum distribution of the leading and subleading jets (the two tagging jets) are shown in Figure~\ref{fig:ptj}. We observe that the NLO and NNLO corrections change the shape of both distributions, being positive or negligible for low momenta and negative for medium or large transverse momentum, where the scale bands completely overlap.

The magnitude of the NNLO corrections is in general moderate, and very rarely exceeds 5\%, while the NLO corrections can be as large as 30\% and leads to a substantial modification of the shape of the distributions. 
Despite partly lying outside the uncertainty bands in the range of the observables we consider, the small magnitude of the NNLO corrections and scale uncertainty bands indicates a good perturbative convergence. 


\begin{table}[t]
    \footnotesize
    \centering
    \begin{subtable}{0.45\textwidth}
        \begin{tabular}{ @{\enskip}l c c@{\enskip} }
            \toprule
                & $\sigma^{\text{reference}}$ (fb) & $\sigma^{\NNLOJET}$ (fb) \\
            \midrule
            LO   & $ 4032^{+57}_{-69} $             & $4032^{+56}_{-69} $ \\
            NLO  & $ 3929^{+24}_{-23} $             & $3927^{+25}_{-24} $ \\
            NNLO & $ 3888^{+16}_{-12} $             & $3884^{+16}_{-12} $ \\
            \bottomrule
        \end{tabular}
        \caption{Fully inclusive VBF cross section. \label{table:fully_inclusive}}
    \end{subtable}\hfill
    \begin{subtable}{0.45\textwidth}
  \centering
  \begin{tabular}{ @{\enskip}l c c@{\enskip} }
    \toprule
         & $\sigma^{\text{reference}}$ (fb) & $\sigma^{\NNLOJET}$ (fb) \\
    \midrule
    LO   &  $ 957^{+66}_{-59} $  & $ 957^{+66}_{-59} $ \\
    NLO  &  $ 876^{+ 8}_{-18} $  & $ 877^{+ 7}_{-17} $ \\
    NNLO &  $ 844^{+ 8}_{- 8} $  & $ 844^{+ 9}_{- 9} $ \\
    \bottomrule
  \end{tabular}
  \caption{Cross section after VBF cuts are applied. \label{table:fiducial}}
    \end{subtable}
    \caption{Comparison with revised results of~\cite{Cacciari:2015jma} for the total cross section. The uncertainty corresponds to a scale variation of $\mu_F = \mu_R = \left\{\frac{1}{2}, 1, 2\right\}\times\mu_0$. $\mu_0$ is given in Eq.~\eqref{eq:scale_choice}. }
\end{table}

\section{Conclusions}
We computed the second-order QCD corrections to Higgs boson production through Vector Boson Fusion in the DIS approximation which has been proven to be a very reliable approximation when VBF cuts are applied.
Our results are implemented in the \NNLOJET framework and fully validated with both our internal test suite and external tools and can be used to compute any infrared-safe observable derived from the VBF process up to \order{\alphas^2}.

In this talk we have summarised the suite of tests that each process in \NNLOJET~\cite{Gehrmann:2018szu} is subjected to and the results published in~\cite{Cruz-Martinez:2018rod}. We also found very good agreement with external tools and independent calculations of the same process.

We observe a kinematical dependence of the NNLO corrections in the distributions of the two leading jets (tagging jets) in transverse momentum, usually amounting to no more than a $5\%$ correction.
Since it is precisely through cuts on these jets that the VBF process is defined, the NNLO effects may have an important impact on the precise efficiency of the VBF cuts, and consequently on all future precision studies of VBF Higgs boson production. 

\bibliographystyle{JHEP}
\newcommand{\url}[1]{{}}
{\footnotesize
    \bibliography{library}
}

\end{document}